# Autonomous detection of molecular configurations in microscopic images based on deep convolutional neural network

Ze-Bin Wu[*]

Condensed Matter Physics and Material Sciences, Brookhaven National Laboratory, Upton 11973, Unites States

Beijing National Laboratory for Condensed Matter Physics and Institute of Physics, Chinese Academy of Sciences, Beijing 100190, China

[*]Corresponding to: zbwu.china@gmail.com

**ABSTRACT:** In an effort to explore high-throughput processing of microscopic image data, a method based on deep convolutional neural network is proposed. The state-of-the-art computer vision algorithm, Faster R-CNN, was trained for the detection of iron (II) phthalocyanines on Se-terminated Au(111) platform resolved by scanning probe microscopy. The construction of the feature pyramid enables the multi-scale molecule detection in images of different scales from 10 nm to 50 nm. After the detection, the orientation of each molecule is measured by a following program. Based on the statistical distribution of the orientation angles, the preferred adsorption configurations of iron (II) phthalocyanine on the platform are revealed. This method yields high accuracy and recall with F1 score close to 1 after optimization of hyperparameters and training. It is expected to be a feasible solution in the scenarios where autonomous and high-throughput processing of microscopic image data is needed.

# 1. INTRODUCTION

Information extraction from microscopic image data requires careful analysis, which relies heavily on human vision. Though it's feasible when the quantity of image data is small, the problems manifest when the quantity becomes big. Not only is the analysis of large quantities of image data a repetitious and laborious job for human, but also many quatitative information is not straightforward to be obtained efficiently. Therefore, autonomous method based on machine vision is highly desirable for the high-throughput processing of microscopic image data.

Machine vision has made remarkable progress with the development of deep learning and convolutional neural network (CNN).[1] CNN enables parameter sharing so that the quantity of parameters is reduced greatly compared with fully-connected neural networks. Besides, the convolutional kernels are able to capture the key features of spatial patterns in images. Therefore, deep learning algorithms based on CNN show good performance in object detection,[2-14] and have wide appilications in industries, for example, in automous driving. Some work on microscopy analysis based on deep neural networks has been reported in recent years, such as the identification of atoms in scanning transmission electron microscopy (STEM) images,[15] the classification of molecular conformational states in scanning probe microscopy (SPM) images,[16] the tracking of defects in STEM movies,[17] and etc.[18-20] These results manifest the great potential of deep neural networks for the processing of microscopic images.

In this paper, an autonomous processing method based on CNN is proposed for the high-throughput processing of microscopic image data. The feasibility of this method is validated on the detection of molecular adsorption configurations in SPM images. Figure 1 shows two SPM images of different scales. FePc molecules adsorb on the single-atomic-layer Se platform, while Se atoms adsorb at the $\sqrt{3} \times \sqrt{3}$ positions on Au(111) substrate. To understand the adsorption behaviors of FePc, statistics based on the orientation information of large quantities of molecules is needed. The state-of-the-art CNN algorithm, known as Faster R-CNN,[5] was trained for the detection of FePc in the SPM images. The detection result has high accuracy and recall. After the detection, a program is used for the measurement of the orientation angle of each molecule. Based on the statistics of the orientation angles, the adsorption behaviors of FePc on this platform is revealed.

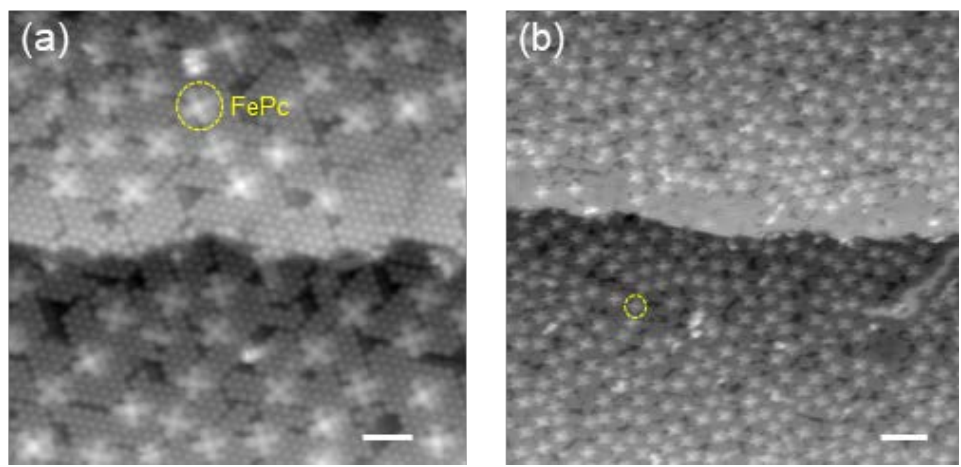

**Figure 1.** Two SPM images of different scales. (a) 18.5 × 18.5 nm$^2$ (the scale bar is 3 nm). (b) 50 × 50 nm$^2$ (the scale bar is 5 nm). The the cross-shape FePc molecules adsorb on Se atomic layers on Au(111) substrate. The orientation of FePc is resolved clearly in the SPM images, which is a prerequisite for the autonomous analysis.

## 2. Methods

There are mainly two different types of CNNs for object detection, single-shot method and multiple-shot method. Single-shot methods, like YOLO (You Only Look Once)[6-8] and SSD (Single Shot MultiBox Detector),[9] have a single forward pass of CNN. While multiple-shot methods, like R-CNNs (Region-based CNNs),[3-5] contain more than one single forward CNNs and relies on the input of region proposals. In general, multiple-shot methods have better performance but are less time-efficient compared with single-shot methods. Given the fact that the acquisition time of one SPM image is at the timescale of minutes or rarely, seconds for fast scanning, which is much longer than the image processing time by either single-shot or multi-shot methods, the multi-shot method is adopted in this project for better performance. Nevertheless, the cutting-edge single-shot method, YOLOv3,[8] has been tried. It shows an unstable outcome for images of different scales. Besides, it has a low recall that many molecules in the image are not detected. While increasing the recall by tuning the hyperparatmters yields a low accuracy.

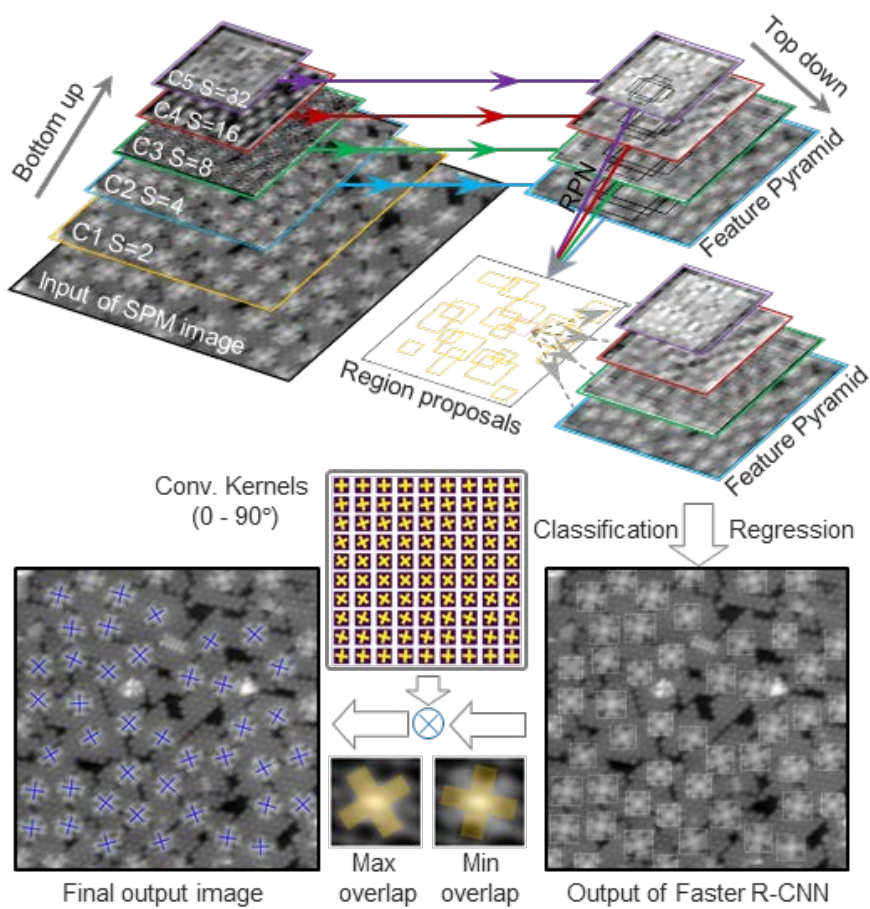

**Figure 2** Schematics of the workflow for molecules detection and orientation measurement, based on Faster R-CNN. "S" means "stride" between adjacent features in each feature map with respect to the raw image ($512^2$ in dimension). C1-C5 indicates convolutional layers at different output levels in ResNet-50. Here, C2-C5 layers are passed to the top-down phase. The colors of the lines indicate the feature maps with the same dimensions and the arrows indicate the data flow of the feature maps. RPN proposes regions according to the feature pyramid. The positions of the molecules are detected by further classification and bounding box regression with respect to the region proposals. The orientation angle of FePc in the region detected by Faster R-CNN is identified by searching for the maximum convolutional sum between that region and each element in a kernel set composed of cross-shape convolutional kernels ranging continuously from 0° to 90°. At last, a mask indicating the orientation is drawn on top of each FePc in the images by the program.

Faster R-CNN[5] is developed from R-CNN[3] and Fast R-CNN.[4] Compared with its parent architectures, Faster R-CNN incorporates the trainable Region Proposal Network (RPN) for the generation of region proposals. In the algorithm used in this paper, Feature Pyramid Network (FPN)[10] is included for the construction of the feature maps, which enables the multi-scale detection. The backbone for feature extraction is ResNet-50.[21] The workflow for molecule detection is presented in Fig. 2. The pipeline of this architecture in detail is presented in the Supplementary Materials. The size of the image data in both training dataset and testing dataset is the same: $512^2$ (512 × 512 pixels), while the physical scales of the images range from 10 nm$^2$ to 50 nm$^2$.

The generation of the feature pyramid is composed of two phases, the bottom-up phase and the top-down phase, as shown in Fig. 2. In the bottom-up phase (i.e. ResNet-50), the semantic information in the features map becomes richer as the network level goes higher (deeper). On the other hand, the details in the feature map become less and less as the dimension of the feature map decreases. The dilemma of rich semantics and enough details in CNNs is avoided by the introduction of the top-down phase, where the feature map at the higher level (smaller dimension) with richer semantics is up-sampled and then concatenated with the feature map at the lower level (larger dimension) with richer details. The feature map data flow between the two phases in the concatenation process is indicated by the colored lines in Fig. 2. The final feature pyramid consists of four different feature maps, the dimensions of which are $16^2$, $32^2$, $64^2$, and $128^2$ from the top to the bottom.

The region proposals are predicted based on the feature pyramid with respect to the pre-defined anchors at each feature element in the feature maps, as indicated by the black arrow labeled by "RPN" in Fig.2. Thereafter, each region proposal is then mapped to a feature by RoIAlign,[14] as shown by the dashed grey arrow in Fig. 2. At last, the classification network evaluates the probability of the presence of FePc in that region, and the bounding box regression network fine-tunes the region proposals to make them fit the molecules tightly. The feature pyramid is the essence of this architecture, which determines the quality of both the region proposals and the detection results. In the region proposing stage by RPN, the design of the anchors, such as the sizes and height-width ratios, are important hyper-parameters for high-quality region proposals in both training and test stages. Because FePc has C4 rotational symmetry, the height-width ratios are set to close to one here. The sizes of the anchors are set to $1^2$, $2^2$, $4^2$, and $8^2$ for each feature level. In the detection stage, the RoIAlign mapping algorithm is optimized in this way that the feature map with large dimension and fine features at the bottom level of the feature pyramid is responsible for the detection of molecules in large-scale images (molecules are "small"), while the feature map with small dimension and coarse features at the top level is responsible for small-scale images (molecules are "big").

The training of deep neural networks requires a large quantity of annotated data. Typically, to train a deep neural network initialized by random parameters, the amount of annotated image data for training is usually tens of thousands or even more. For this reason, there are many public annotated image datasets available, such as ImageNet,[22] Microsoft COCO,[23] and PASCAL VOC.[24] In the microcopy field, there are not such public image datasets. In this project, by transfer learning,[25] it is verified that only hundreds of annotated SPM images are needed in the training dataset to achieve a fairly good result. However, annotating hundreds of SPM images manually is still a laborious task, which contradicts the intention of this project. This problem is addressed by the label-preserving data augmentation techniques. Those manually-labeled images are split, rotated, mirrored, cropped, and duplicated to increase the quantity of samples in the training dataset. In this way, the human labor involved in the image annotation is greatly reduced. As a result, there are 800 annotated images in the training dataset, while only 6 raw SPM images need to be annotated manually. To train the capability of the multi-scale detection, images of different scales (10 - 50 nm$^2$) are included in the training dataset with an optimized amount ratio.

In the output of the Faster R-CNN, FePc molecules are marked by bounding boxes in the SPM images, as shown in the image at the bottom right in Fig. 2. The orientation angle of each molecule is measured by a Python program which calculates the convolutional sum between the bounded area and each measurement kernel in a kernel set. The kernel set is composed of images with cross-shape features resembling FePc in SPM images. The orientation angles of those measurement kernels range from 0° to 90° with a precision of 1°. The kernel which overlaps most with the molecule results in the maximum convolutional sum. Thus, the orientation angles of the detected molecules are marked by the kernel which yields maximum convolutional sum. At last, the cross-shape masks are labeled on top of the molecules, as shown in the final output image at the bottom left in Fig. 2.

The experiments of this CNN-based workflow were carried out on a normal workstation, equipped with a single GPU (RTX 2080Ti, 11 G graphical memory). The backend of the deep learning architecture is PyTorch. The Faster R-CNN codes with feature pyramid network is available in an open-source object detection toolbox, MMDetection[26], and is used in this project. The images in the training dataset are annotated using Labelme.[27] The LoG algorithm is imported from the image processing Python package - Scikit-image. The FePc/Se/Au(111) sample was prepared and the SPM images were obtained in a home-made cryogenic SPM system.[28]

## 3. Results

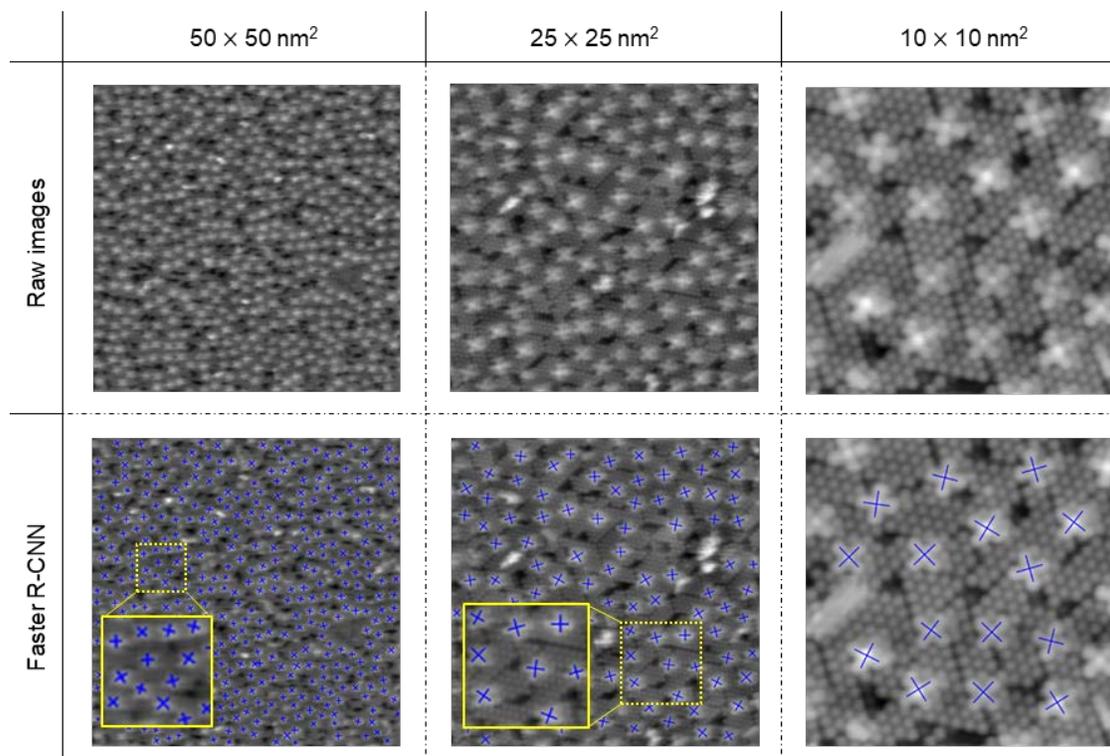

**Figure 3** The first row shows three characteristic raw SPM images of different scales. The second row shows the raw output of the autonomous workflow based on Faster R-CNN. The blue cross lables of FePc are produced by the program autonomously.

All the test images for Faster R-CNN are not included in the training dataset. Some test results are presented in Fig. 3. Images of different scales from 10 × 10 nm$^2$ to 50 × 50 nm$^2$, are randomly selected and shown in the first column of Fig.

3. Molecules in all images of different scales are detected with high accuracy ( > 99% ) and recall ( > 99% ), even in the large-scale images where the objects are small. In the 50 × 50 nm$^2$ and 25 × 25 nm$^2$ images, only molecules are detected by Faster R-CNN, while the defect features are recognized and excluded from the detection results. In the 10 × 10 nm$^2$ image, FePc molecules are detected but blob-like Se atoms are not.

The orientation information of thousands of FePc is resolved in all available SPM images. The distribution histogram of the orientation angles is displayed in Fig. 4a. The orientation angle of FePc is defined with respect to the image direction, as shown in the inset of Fig. 4a. The orientation of FePc is not random on this platform. Considering the quasi C3 rotational symmetry of the Se platform and the C4 rotational symmetry of FePc, molecules with an orientation difference of 30° are categorized as equivalent adsorption configurations. Therefore, orientation angles close to 0°, 30°, 60° are marked in blue in the histogram, indicating configuration Ⅰ, while those close to 15°, 45°, 75° are marked in green, indicating configuration Ⅱ.

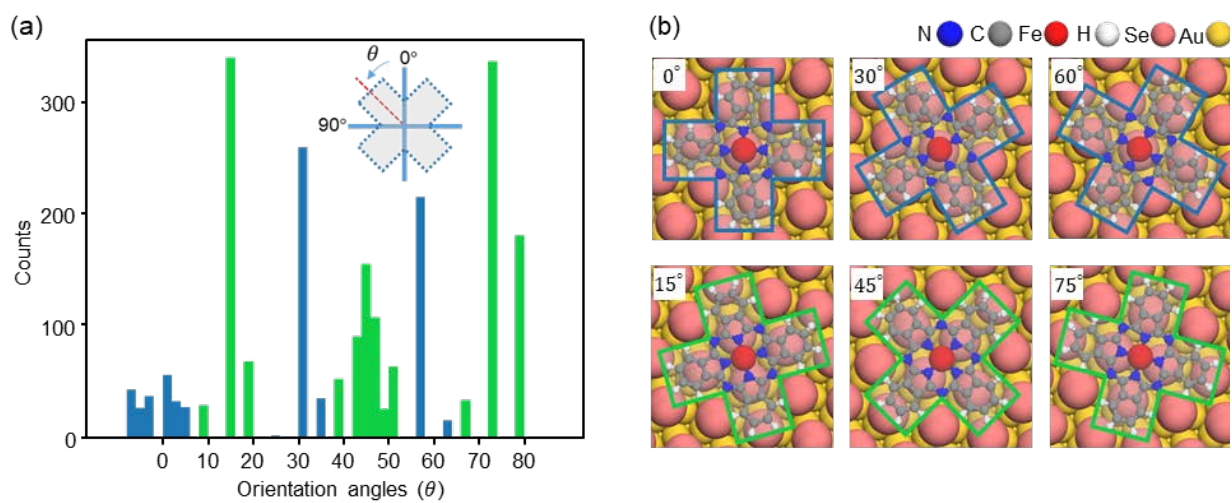

**Figure 4** Adsorption configurations of FePc on Se-terminated Au(111). (a) Statistic histogram on the distribution of FePc orientations based on the output of the method. The orientations can be roughly categorized into two groups, colored by green (~ 15°, 45°, 75°) and blue (~ 0°, 30°, 60°), indicating two dominating adsorption configurations of FePc. (b) Proposed atomic models of the two adsorption configurations. In the first row, Fe atom in FePc adsorbs on top of the Se atom. In the second row, Fe atom adsorbs at the bridge site of two Se atoms.

Metal phthalocyanine (MPc) is an ideal building block for various nanostructures by molecular self-assembly[29-32] The properties of MPc in the nanostructures are modulated by the adsorption configurations.[33,34] In Figure 4b, two proposed atomic models are constructed according to some high-resolution SPM images. In configuration Ⅰ, the central Fe atom of FePc adsorbs on top of Se atom, while in configuration Ⅱ, the central Fe atom adsorbs at the Se-Se bridge site. Therefore, FePc has similar adsorption configurations on this platform as on bare Au(111) substrate,[29] which indicates that these two adsorption configurations can be universal for FePc molecules on platforms or substrates with C3 rotational symmetry. The deviation of the orientation angles from the dominating value can be caused by many reasons, such as the imperfection of the Se platform, reconstruction of the Au(111) surface, etc.

## 4. CONCLUSIONS

A high-throughput processing method based on CNN is proposed and applied to the detection of adsorption configurations of molecules in SPM images. By optimizing the hyper-parameters through experiments, this method yields a high recall and accuracy (F1 score close to 1). Transfer learning makes the training of Faster R-CNN for SPM images detection very efficient. Because the acquisition time of one SPM image is much longer than the processing time, this method can be a promising solution for real-time analysis of streaming microscopic data.

Data availability

The architecture of Faster R-CNN and Python codes for the image analysis can be found in the supplementary materials. The configuration files and well-trained parameters for Faster R-CNN used in this project are available from the author upon request.

**ACKNOWLEDGEMENTS**

The author thanks the MMDetection team from CUHK for realeasing the source codes of their Faster R-CNN architecture, which makes this state-of-the-art method for object detection ready to be applied to microscopy analysis. The author thanks H. J. Gao, G. Li, and R. T. Wu for the discussion.

# REFERENCES

(1) Goodfellow I., Bengio Y. & Courville A. Deep learning. (MIT Press, Cambridge, 2016).
(2) Krizhevsky, A., Sutskever, I. & Hinton, G. E. Imagenet classification with deep convolutional neural networks. In *NIPS*, (2012).
(3) Girshick, R., Donahue, J., Darrell, T. & Malik, J. Rich feature hierarchies for accurate object detection and semantic segmentation. In *CVPR* (2014).
(4) Girshick, R. Fast R-CNN. In *ICCV* (2015).
(5) Ren, S., He, K., Girshick, R. & Sun, J. Faster R-CNN: towards real-time object detection with region proposal networks. In *NIPS* (2015).
(6) Redmon, J., Divvala, S., Girshick, R. & Farhadi, A. You only look once: unified, real-time object detection. In *CVPR* (2016).
(7) Redmon, J. & Farhadi, A. YOLO9000: better, faster, stronger. In *CVPR* (2017).
(8) Redmon, J. & Farhadi, A. YOLOv3: an incremental improvement. In *CVPR* (2018).
(9) Liu, W. et al. SSD: single shot multibox detector. In *ECCV* (2016).
(10) Lin, T. et al. Feature pyramid networks for object detection. In *CVPR* (2017).
(11) Ronneberger, O., Fischer, P. & Brox, T. U-Net: Convolutional networks for biomedical image segmentation. In *MICCAI* (2015).
(12) Long, J., Shelhamer, E. & Darrell, T. Fully convolutional networks for semantic segmentation. In *CVPR* (2015).
(13) Badrinarayanan, V., Kendall, A. & Cipolla, R. SegNet: A deep convolutional encoder-decoder architecture for image segmentation. *PAMI* **39,** 2481–2495 (2017).
(14) He, K., Gkioxari, G., Dollár, P. & Girshick, R. Mask R-CNN. In *ICCV* (2017).
(15) Ziatdinov, M. et al. Deep learning of atomically resolved scanning transmission electron microscopy images: chemical identification and tracking local transformations. *ACS Nano* **11,** 12742-12752 (2017).
(16) Ziatdinov, M., Maksov, A. & Kalinin, S. V. Learning surface molecular structures via machine vision. *npj Comput. Mater.* **3,** 31 (2017).
(17) Maksov, A. et al. Deep learning analysis of defect and phase evolution during electron beam-induced transformations in $WS_2$. *npj Comput. Mater.* **5,** 12 (2019).
(18) Kalinin, S. V., Sumpter, B. G. & Archibald, R. K. Big–deep–smart data in imaging for guiding materials design. *Nat. Mater.* **14,** 973-980 (2015).
(19) Kalinin, S. V. et al. Big, deep, and smart data in scanning probe microscopy. *ACS Nano* **10,** 9068-9086 (2016).
(20) Rashidi, M. et al. Deep learning-guided surface characterization for autonomous hydrogen lithography. *Mach. Learn.: Sci. Technol.* **1,** 025001 (2020).
(21) He, K., Zhang, X., Ren, S. & Sun, J. Deep residual learning for image recognition. In *CVPR* (2016).
(22) Russakovsky, O. et al. ImageNet large scale visual recognition challenge. *Int. J. Comput. Vis.* **115,** 211–252 (2015).
(23) Lin, T. Y. et al. Microsoft COCO: Common objects in context. Preprint at https://arxiv.org/pdf/1405.0312.pdf (2015)
(24) Everingham, M., Van Gool, L., Williams, C. K. I., Winn, J. & Zisserman, A. The PASCAL visual object classes (VOC) challenge. *Int. J. Comput. Vis.* **88,** 303–338 (2010).
(25) Weiss, K., Khoshgoftaar, T. M. & Wang, D. A survey of transfer learning. *J. Big Data* **3,** 9 (2016).
(26) Chen, K. et al. MMDetection: Open MMLab detection toolbox and benchmark. Preprint at https://arxiv.org/pdf/1906.07155.pdf (2019).
(27) Wada, K. Labelme: image polygonal annotation with Python. *https://github.com/wkentaro/labelme* (2016).
(28) Wu, Z. B. *et al*. A low-temperature scanning probe microscopy system with molecular beam epitaxy and optical access. *Rev. Sci. Instrum.* **89,** 113705 (2018).
(29) Gao, L. et al. Site-specific Kondo effect at ambient temperatures in iron-based molecules. *Phys. Rev. Lett.* **99,** 106402 (2007).
(30) Girovsky, J. et al. Long-range ferrimagnetic order in a two-dimensional supramolecular Kondo lattice. *Nat. Comm.* **8,** 15388 (2017).
(31) Hiraoka, R. et al. Single-molecule quantum dot as a Kondo simulator. *Nat. Comm.* **8,** 16012 (2017).

(32) Sorokin, A. B. Phthalocyanine metal complexes in catalysis. *Chem. Rev.* **113,** 8152-8191 (2013).
(33) Tsukahara, N. et al. Adsorption-induced switching of magnetic anisotropy in a single Iron(II) phthalocyanine molecule on an oxidized Cu(110) surface. *Phys. Rev. Lett.* **102,** 167203 (2009).
(34) Minamitani, E. et al. M. Symmetry-driven novel Kondo effect in a molecule. Phys. Rev. Lett. 109, 086602 (2012).

**Supplementary materials**

Jupyter Notebook for

Autonomous detection of molecular configurations in microscopic images

based on deep convolutional neural network

Ze-Bin Wu[*]

Condensed Matter Physics and Material Sciences, Brookhaven National Laboratory, Upton 11973, Unites States

Beijing National Laboratory for Condensed Matter Physics and Institute of Physics, Chinese Academy of Sciences, Beijing 100190, China

[*]Corresponding to: zbwu.china@gmail.com

```python
from mmdet.apis import init_detector, inference_detector, show_result
from PIL import Image
from pylab import *
import cv2
import matplotlib as mpl
import glob2
import os
from matplotlib.pyplot import *
from numpy import *
from scipy.stats import multivariate_normal
from scipy.ndimage import rotate
import csv
from scipy.ndimage import filters
```

### Loading CNN model and initialization with trained parameters

```python
config_file = 'configs/round#6/faster_rcnn_r50_fpn_1x_cus04.py'
checkpoint_file = 'work_dirs/faster_rcnn_r50_fpn_1x_cus04/latest.pth'
model = init_detector(config_file, checkpoint_file, device='cuda:0')
```

### Load image data from test dataset

```python
img_pths = glob2.glob("test_set/other/*.png")
```

### Detection result on one image

```python
figure(figsize=(20,20))
result = inference_detector(model, 'test_set/other/Pt_572.grey.png')
im = Image.open('test_set/other/Pt_572.grey.png')
im = array(im)
imshow(im)
gray()
for i, box in enumerate(result[0]):
    if i == 51 or i == 52: continue
    
    plot([box[0], box[2]],[box[1], box[1]], 'w-.')
    plot([box[0], box[2]], [box[3] , box[3]], 'w-.')
    plot([box[0], box[0]],[box[1], box[3]], 'w-.')
    plot([box[2], box[2]], [box[1] , box[3]], 'w-.')
axis('off')
savefig("test_set/other/result/572_bbox.png", bbox_inches='tight', pad_inches = 0)
```

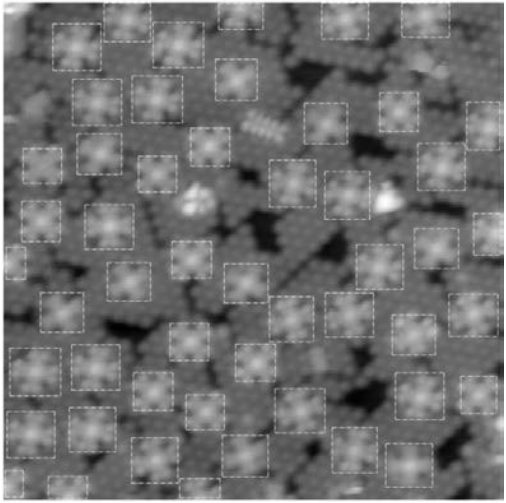

**Detection results on more test images**

```
figure(figsize = (100,80))
for i, img_pth in enumerate(img_pths):
    result = inference_detector(model, img_pth)
    im = Image.open(img_pth)
    im = array(im)
    subplot(4,5,i+1)
    imshow(im)
    gray()
    for box in result[0]:
        plot([box[0], box[2]],[box[1], box[1]], 'w-.')
        plot([box[0], box[2]], [box[3] , box[3]], 'w-.')
        plot([box[0], box[0]],[box[1], box[3]], 'w-.')
        plot([box[2], box[2]], [box[1] , box[3]], 'w-.')
    axis('off')
```

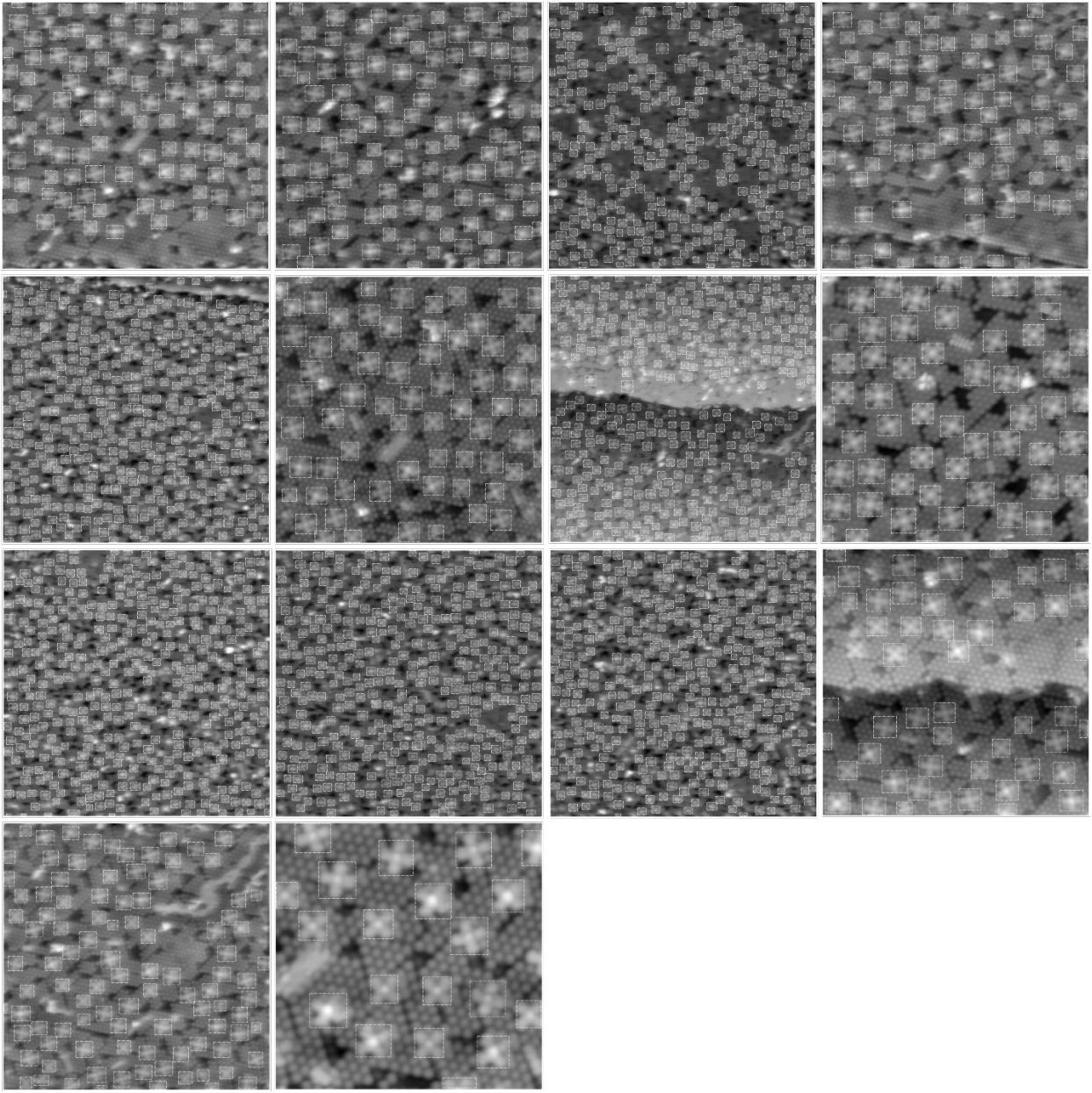

**Construction of angle measurement kernel set**

```
x_1 = [1] * 56
x_2 = [0] * ((256 - len(x_1))//2)
x = x_2 + x_1 + x_2
x = array(x)
x = x.reshape(1,256)
x = repeat(x,256,axis=0)
filt_flat_tmp = uint8(x + x.T)
filt_flat_tmp[len(x_2):len(x_2)+len(x_1),len(x_2):len(x_2)+len(x_1)] = 1
filt_flat_orig = zeros((256,256), dtype = int)
space_cut = 20
filt_flat_orig[space_cut:-space_cut,space_cut:-space_cut] = filt_flat_tmp[space_cut:-space_cut,space_cut:-space_cut]
filts = []
figure(figsize=(10,10))
for i in range(90):
    filt_tmp = rotate(filt_flat_orig, i)
    ctr = len(filt_tmp)//2
    filt = filt_tmp[ctr-128:ctr+128, ctr-128:ctr+128]
    filts.append(filt)
    subplot(10,9,i+1)
    axis('off')
    imshow(filt)
```

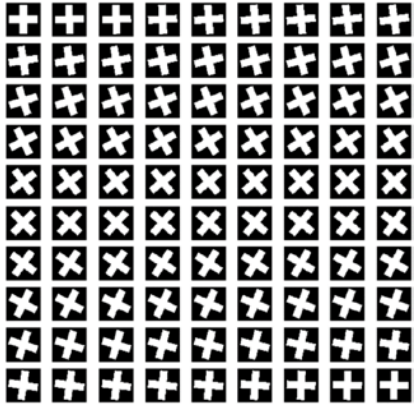

**Construction of orientation angle measurement function**

```
def rotation_detector(img, box, kernels):
    ctr_x = int(ceil((box[0] + box[2])//2))
    ctr_y = int(ceil((box[1] + box[3])//2))
    scale = int(ceil(sqrt((box[2]-box[0])*(box[3]-box[1]))))
    mol = None
if (ctr_y-scale//2) >=0 and (ctr_x-scale//2)>=0 and (ctr_y+scale//2)<=img.shape[0] and (ctr_x+scale//2)<=img.shape[1]:
    if scale%2 == 0 :
        mol = img[(ctr_y-scale//2):(ctr_y+scale//2), (ctr_x-scale//2):(ctr_x+scale//2)]
        mol = cv2.resize(mol, None, fx = 256/scale, fy = 256/scale, interpolation = cv2.INTER_LINEAR_EXACT)
    elif scale%2 != 0 and (ctr_y+scale//2+1)<=img.shape[0] and (ctr_x+scale//2+1)<=img.shape[1]:
        mol = img[(ctr_y-scale//2):(ctr_y+1+scale//2), (ctr_x-scale//2):(ctr_x+1+scale//2)]
        mol = cv2.resize(mol, None, fx = 256/scale, fy = 256/scale, interpolation = cv2.INTER_LINEAR_EXACT)
if mol is None:
    return None, None, None, None
val = []
for i, kernel in enumerate(kernels):
    assert(kernel.shape == mol.shape)
    val.append((kernel * mol).sum())
angle = val.index(max(val))
return ctr_x, ctr_y, scale, angle
```

**Measurement of orientation angles and labeling the molecules with cross-shape masks**

```
orientations = []
figure(figsize=(80,80))
subplots_adjust(left=0, right=1, bottom=0, top=1, hspace=0.05, wspace=0.05)
mpl.rcParams['lines.linewidth'] = 4
for i, img_pth in enumerate(img_pths):
    orientations.append([])
    img = Image.open(img_pth)
    img = array(img)
    #print(im_559_01.shape)
    img = cv2.resize(img, None, fx = 512/img.shape[1], fy = 512/img.shape[0], interpolation = cv2.INTER_LINEAR_EXACT)
    subplot(4,4,i+1)
    imshow(img)
    gray()
    result = inference_detector(model, img_pth)
    for j,box in enumerate(result[0]):
        if img_pth == 'test_set/other/Pt_572.grey.png':
            if j == 51 or j == 52: continue
        ctr_x, ctr_y, scale, angle = rotation_detector(img[:,:,0], box, filts)
        if ctr_x is None: continue
        orientations[i].append(angle)
        scale *= 0.6
        rot_deg = pi * (angle/180)
        if rot_deg <= pi/4:
            plot([ctr_x,ctr_x+scale//2*tan(rot_deg)],[ctr_y,ctr_y+scale//2],'b') # 4th quadro
            plot([ctr_x,ctr_x-scale//2*tan(rot_deg)],[ctr_y,ctr_y-scale//2],'b') # 1st quadro
            plot([ctr_x,ctr_x+scale//2],[ctr_y,ctr_y-scale//2*tan(rot_deg)],'b') # 2nd quadro
            plot([ctr_x,ctr_x-scale//2],[ctr_y,ctr_y+scale//2*tan(rot_deg)],'b') #3rd quadro
        if rot_deg > pi/4:
            plot([ctr_x,ctr_x+scale//2],[ctr_y,ctr_y+scale//(2*tan(rot_deg))],'b') # 4th quadro
            plot([ctr_x,ctr_x-scale//2],[ctr_y,ctr_y-scale//(2*tan(rot_deg))],'b') # 1st quadro
            plot([ctr_x,ctr_x+scale//(2*tan(rot_deg))],[ctr_y,ctr_y-scale//2],'b') # 2nd quadro
            plot([ctr_x,ctr_x-scale//(2*tan(rot_deg))],[ctr_y,ctr_y+scale//2],'b') #3rd quadro
```

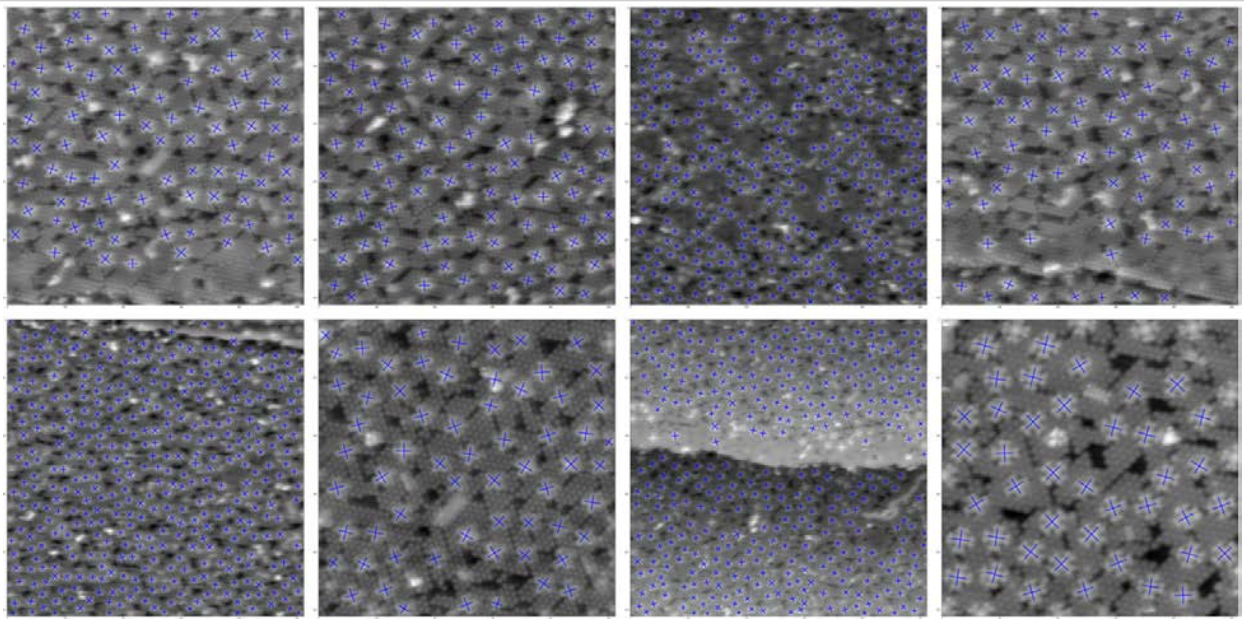

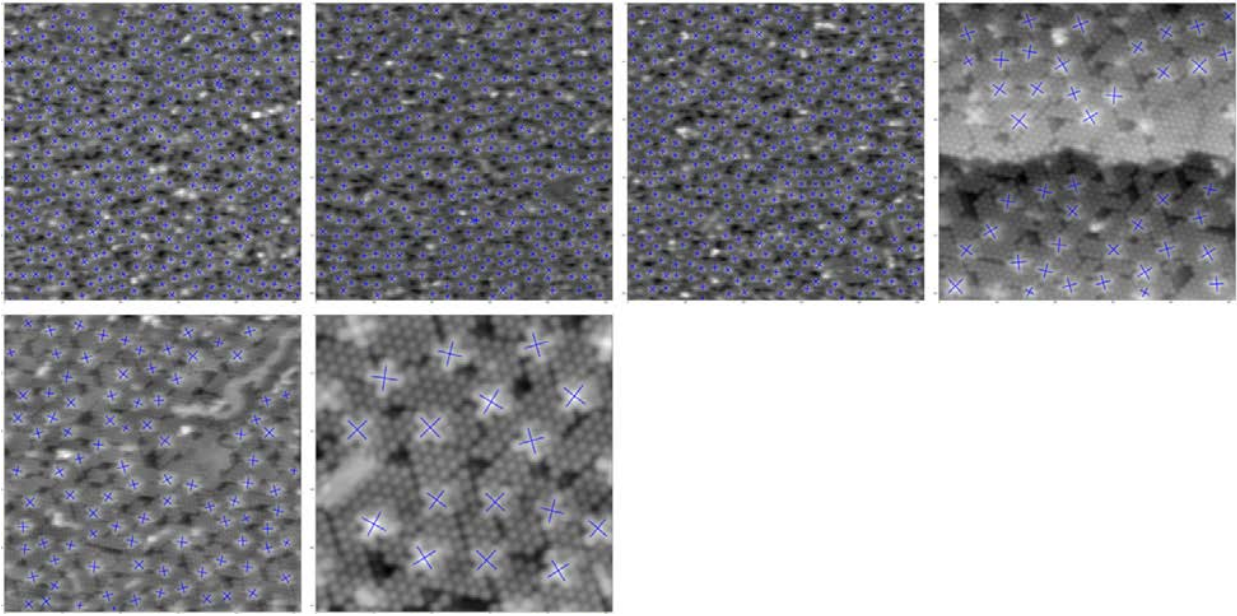

## Modules in this Faster RCNN architecture  (More details in _https://arxiv.org/pdf/1906.07155.pdf (2019)_)

```
model.modules
```

<bound method Module.modules of FasterRCNN(
  **(backbone): ResNet(**
    (conv1): Conv2d(3, 64, kernel_size=(7, 7), stride=(2, 2), padding=(3, 3), bias=False)
    (bn1): BatchNorm2d(64, eps=1e-05, momentum=0.1, affine=True, track_running_stats=True)
    (relu): ReLU(inplace=True)
    (maxpool): MaxPool2d(kernel_size=3, stride=2, padding=1, dilation=1, ceil_mode=False)
    (layer1): Sequential(
      (0): Bottleneck(
        (conv1): Conv2d(64, 64, kernel_size=(1, 1), stride=(1, 1), bias=False)
        (bn1): BatchNorm2d(64, eps=1e-05, momentum=0.1, affine=True, track_running_stats=True)
        (conv2): Conv2d(64, 64, kernel_size=(3, 3), stride=(1, 1), padding=(1, 1), bias=False)
        (bn2): BatchNorm2d(64, eps=1e-05, momentum=0.1, affine=True, track_running_stats=True)
        (conv3): Conv2d(64, 256, kernel_size=(1, 1), stride=(1, 1), bias=False)
        (bn3): BatchNorm2d(256, eps=1e-05, momentum=0.1, affine=True, track_running_stats=True)
        (relu): ReLU(inplace=True)
        (downsample): Sequential(
          (0): Conv2d(64, 256, kernel_size=(1, 1), stride=(1, 1), bias=False)
          (1): BatchNorm2d(256, eps=1e-05, momentum=0.1, affine=True, track_running_stats=True)
        )
      )
      (1): Bottleneck(
        (conv1): Conv2d(256, 64, kernel_size=(1, 1), stride=(1, 1), bias=False)
        (bn1): BatchNorm2d(64, eps=1e-05, momentum=0.1, affine=True, track_running_stats=True)
        (conv2): Conv2d(64, 64, kernel_size=(3, 3), stride=(1, 1), padding=(1, 1), bias=False)
        (bn2): BatchNorm2d(64, eps=1e-05, momentum=0.1, affine=True, track_running_stats=True)
        (conv3): Conv2d(64, 256, kernel_size=(1, 1), stride=(1, 1), bias=False)
        (bn3): BatchNorm2d(256, eps=1e-05, momentum=0.1, affine=True, track_running_stats=True)
        (relu): ReLU(inplace=True)
      )
      (2): Bottleneck(
        (conv1): Conv2d(256, 64, kernel_size=(1, 1), stride=(1, 1), bias=False)
        (bn1): BatchNorm2d(64, eps=1e-05, momentum=0.1, affine=True, track_running_stats=True)
        (conv2): Conv2d(64, 64, kernel_size=(3, 3), stride=(1, 1), padding=(1, 1), bias=False)
        (bn2): BatchNorm2d(64, eps=1e-05, momentum=0.1, affine=True, track_running_stats=True)
        (conv3): Conv2d(64, 256, kernel_size=(1, 1), stride=(1, 1), bias=False)
        (bn3): BatchNorm2d(256, eps=1e-05, momentum=0.1, affine=True, track_running_stats=True)
        (relu): ReLU(inplace=True)
      )
    )
  )

```
    (layer2): Sequential(
      (0): Bottleneck(
        (conv1): Conv2d(256, 128, kernel_size=(1, 1), stride=(1, 1), bias=False)
        (bn1): BatchNorm2d(128, eps=1e-05, momentum=0.1, affine=True, track_running_stats=True)
        (conv2): Conv2d(128, 128, kernel_size=(3, 3), stride=(2, 2), padding=(1, 1), bias=False)
        (bn2): BatchNorm2d(128, eps=1e-05, momentum=0.1, affine=True, track_running_stats=True)
        (conv3): Conv2d(128, 512, kernel_size=(1, 1), stride=(1, 1), bias=False)
        (bn3): BatchNorm2d(512, eps=1e-05, momentum=0.1, affine=True, track_running_stats=True)
        (relu): ReLU(inplace=True)
        (downsample): Sequential(
          (0): Conv2d(256, 512, kernel_size=(1, 1), stride=(2, 2), bias=False)
          (1): BatchNorm2d(512, eps=1e-05, momentum=0.1, affine=True, track_running_stats=True)
        )
      )
      (1): Bottleneck(
        (conv1): Conv2d(512, 128, kernel_size=(1, 1), stride=(1, 1), bias=False)
        (bn1): BatchNorm2d(128, eps=1e-05, momentum=0.1, affine=True, track_running_stats=True)
        (conv2): Conv2d(128, 128, kernel_size=(3, 3), stride=(1, 1), padding=(1, 1), bias=False)
        (bn2): BatchNorm2d(128, eps=1e-05, momentum=0.1, affine=True, track_running_stats=True)
        (conv3): Conv2d(128, 512, kernel_size=(1, 1), stride=(1, 1), bias=False)
        (bn3): BatchNorm2d(512, eps=1e-05, momentum=0.1, affine=True, track_running_stats=True)
        (relu): ReLU(inplace=True)
      )
      (2): Bottleneck(
        (conv1): Conv2d(512, 128, kernel_size=(1, 1), stride=(1, 1), bias=False)
        (bn1): BatchNorm2d(128, eps=1e-05, momentum=0.1, affine=True, track_running_stats=True)
        (conv2): Conv2d(128, 128, kernel_size=(3, 3), stride=(1, 1), padding=(1, 1), bias=False)
        (bn2): BatchNorm2d(128, eps=1e-05, momentum=0.1, affine=True, track_running_stats=True)
        (conv3): Conv2d(128, 512, kernel_size=(1, 1), stride=(1, 1), bias=False)
        (bn3): BatchNorm2d(512, eps=1e-05, momentum=0.1, affine=True, track_running_stats=True)
        (relu): ReLU(inplace=True)
      )
      (3): Bottleneck(
        (conv1): Conv2d(512, 128, kernel_size=(1, 1), stride=(1, 1), bias=False)
        (bn1): BatchNorm2d(128, eps=1e-05, momentum=0.1, affine=True, track_running_stats=True)
        (conv2): Conv2d(128, 128, kernel_size=(3, 3), stride=(1, 1), padding=(1, 1), bias=False)
        (bn2): BatchNorm2d(128, eps=1e-05, momentum=0.1, affine=True, track_running_stats=True)
        (conv3): Conv2d(128, 512, kernel_size=(1, 1), stride=(1, 1), bias=False)
        (bn3): BatchNorm2d(512, eps=1e-05, momentum=0.1, affine=True, track_running_stats=True)
        (relu): ReLU(inplace=True)
      )
    )
    (layer3): Sequential(
      (0): Bottleneck(
        (conv1): Conv2d(512, 256, kernel_size=(1, 1), stride=(1, 1), bias=False)
        (bn1): BatchNorm2d(256, eps=1e-05, momentum=0.1, affine=True, track_running_stats=True)
        (conv2): Conv2d(256, 256, kernel_size=(3, 3), stride=(2, 2), padding=(1, 1), bias=False)
        (bn2): BatchNorm2d(256, eps=1e-05, momentum=0.1, affine=True, track_running_stats=True)
        (conv3): Conv2d(256, 1024, kernel_size=(1, 1), stride=(1, 1), bias=False)
        (bn3): BatchNorm2d(1024, eps=1e-05, momentum=0.1, affine=True, track_running_stats=True)
        (relu): ReLU(inplace=True)
        (downsample): Sequential(
          (0): Conv2d(512, 1024, kernel_size=(1, 1), stride=(2, 2), bias=False)
          (1): BatchNorm2d(1024, eps=1e-05, momentum=0.1, affine=True, track_running_stats=True)
        )
      )
      (1): Bottleneck(
        (conv1): Conv2d(1024, 256, kernel_size=(1, 1), stride=(1, 1), bias=False)
        (bn1): BatchNorm2d(256, eps=1e-05, momentum=0.1, affine=True, track_running_stats=True)
        (conv2): Conv2d(256, 256, kernel_size=(3, 3), stride=(1, 1), padding=(1, 1), bias=False)
        (bn2): BatchNorm2d(256, eps=1e-05, momentum=0.1, affine=True, track_running_stats=True)
        (conv3): Conv2d(256, 1024, kernel_size=(1, 1), stride=(1, 1), bias=False)
        (bn3): BatchNorm2d(1024, eps=1e-05, momentum=0.1, affine=True, track_running_stats=True)
        (relu): ReLU(inplace=True)
      )
      (2): Bottleneck(
        (conv1): Conv2d(1024, 256, kernel_size=(1, 1), stride=(1, 1), bias=False)
        (bn1): BatchNorm2d(256, eps=1e-05, momentum=0.1, affine=True, track_running_stats=True)
        (conv2): Conv2d(256, 256, kernel_size=(3, 3), stride=(1, 1), padding=(1, 1), bias=False)
        (bn2): BatchNorm2d(256, eps=1e-05, momentum=0.1, affine=True, track_running_stats=True)
        (conv3): Conv2d(256, 1024, kernel_size=(1, 1), stride=(1, 1), bias=False)
        (bn3): BatchNorm2d(1024, eps=1e-05, momentum=0.1, affine=True, track_running_stats=True)
        (relu): ReLU(inplace=True)
      )
      (3): Bottleneck(
        (conv1): Conv2d(1024, 256, kernel_size=(1, 1), stride=(1, 1), bias=False)
        (bn1): BatchNorm2d(256, eps=1e-05, momentum=0.1, affine=True, track_running_stats=True)
        (conv2): Conv2d(256, 256, kernel_size=(3, 3), stride=(1, 1), padding=(1, 1), bias=False)
        (bn2): BatchNorm2d(256, eps=1e-05, momentum=0.1, affine=True, track_running_stats=True)
        (conv3): Conv2d(256, 1024, kernel_size=(1, 1), stride=(1, 1), bias=False)
```

```
      (bn3): BatchNorm2d(1024, eps=1e-05, momentum=0.1, affine=True, track_running_stats=True)
      (relu): ReLU(inplace=True)
    )
    (4): Bottleneck(
      (conv1): Conv2d(1024, 256, kernel_size=(1, 1), stride=(1, 1), bias=False)
      (bn1): BatchNorm2d(256, eps=1e-05, momentum=0.1, affine=True, track_running_stats=True)
      (conv2): Conv2d(256, 256, kernel_size=(3, 3), stride=(1, 1), padding=(1, 1), bias=False)
      (bn2): BatchNorm2d(256, eps=1e-05, momentum=0.1, affine=True, track_running_stats=True)
      (conv3): Conv2d(256, 1024, kernel_size=(1, 1), stride=(1, 1), bias=False)
      (bn3): BatchNorm2d(1024, eps=1e-05, momentum=0.1, affine=True, track_running_stats=True)
      (relu): ReLU(inplace=True)
    )
    (5): Bottleneck(
      (conv1): Conv2d(1024, 256, kernel_size=(1, 1), stride=(1, 1), bias=False)
      (bn1): BatchNorm2d(256, eps=1e-05, momentum=0.1, affine=True, track_running_stats=True)
      (conv2): Conv2d(256, 256, kernel_size=(3, 3), stride=(1, 1), padding=(1, 1), bias=False)
      (bn2): BatchNorm2d(256, eps=1e-05, momentum=0.1, affine=True, track_running_stats=True)
      (conv3): Conv2d(256, 1024, kernel_size=(1, 1), stride=(1, 1), bias=False)
      (bn3): BatchNorm2d(1024, eps=1e-05, momentum=0.1, affine=True, track_running_stats=True)
      (relu): ReLU(inplace=True)
    )
  )
  (layer4): Sequential(
    (0): Bottleneck(
      (conv1): Conv2d(1024, 512, kernel_size=(1, 1), stride=(1, 1), bias=False)
      (bn1): BatchNorm2d(512, eps=1e-05, momentum=0.1, affine=True, track_running_stats=True)
      (conv2): Conv2d(512, 512, kernel_size=(3, 3), stride=(2, 2), padding=(1, 1), bias=False)
      (bn2): BatchNorm2d(512, eps=1e-05, momentum=0.1, affine=True, track_running_stats=True)
      (conv3): Conv2d(512, 2048, kernel_size=(1, 1), stride=(1, 1), bias=False)
      (bn3): BatchNorm2d(2048, eps=1e-05, momentum=0.1, affine=True, track_running_stats=True)
      (relu): ReLU(inplace=True)
      (downsample): Sequential(
        (0): Conv2d(1024, 2048, kernel_size=(1, 1), stride=(2, 2), bias=False)
        (1): BatchNorm2d(2048, eps=1e-05, momentum=0.1, affine=True, track_running_stats=True)
      )
    )
    (1): Bottleneck(
      (conv1): Conv2d(2048, 512, kernel_size=(1, 1), stride=(1, 1), bias=False)
      (bn1): BatchNorm2d(512, eps=1e-05, momentum=0.1, affine=True, track_running_stats=True)
      (conv2): Conv2d(512, 512, kernel_size=(3, 3), stride=(1, 1), padding=(1, 1), bias=False)
      (bn2): BatchNorm2d(512, eps=1e-05, momentum=0.1, affine=True, track_running_stats=True)
      (conv3): Conv2d(512, 2048, kernel_size=(1, 1), stride=(1, 1), bias=False)
      (bn3): BatchNorm2d(2048, eps=1e-05, momentum=0.1, affine=True, track_running_stats=True)
      (relu): ReLU(inplace=True)
    )
    (2): Bottleneck(
      (conv1): Conv2d(2048, 512, kernel_size=(1, 1), stride=(1, 1), bias=False)
      (bn1): BatchNorm2d(512, eps=1e-05, momentum=0.1, affine=True, track_running_stats=True)
      (conv2): Conv2d(512, 512, kernel_size=(3, 3), stride=(1, 1), padding=(1, 1), bias=False)
      (bn2): BatchNorm2d(512, eps=1e-05, momentum=0.1, affine=True, track_running_stats=True)
      (conv3): Conv2d(512, 2048, kernel_size=(1, 1), stride=(1, 1), bias=False)
      (bn3): BatchNorm2d(2048, eps=1e-05, momentum=0.1, affine=True, track_running_stats=True)
      (relu): ReLU(inplace=True)
    )
  )
)
(neck): FPN(
  (lateral_convs): ModuleList(
    (0): ConvModule(
      (conv): Conv2d(256, 256, kernel_size=(1, 1), stride=(1, 1))
    )
    (1): ConvModule(
      (conv): Conv2d(512, 256, kernel_size=(1, 1), stride=(1, 1))
    )
    (2): ConvModule(
      (conv): Conv2d(1024, 256, kernel_size=(1, 1), stride=(1, 1))
    )
    (3): ConvModule(
      (conv): Conv2d(2048, 256, kernel_size=(1, 1), stride=(1, 1))
    )
  )
```

```
      (fpn_convs): ModuleList(
        (0): ConvModule(
          (conv): Conv2d(256, 256, kernel_size=(3, 3), stride=(1, 1), padding=(1, 1))
        )
        (1): ConvModule(
          (conv): Conv2d(256, 256, kernel_size=(3, 3), stride=(1, 1), padding=(1, 1))
        )
        (2): ConvModule(
          (conv): Conv2d(256, 256, kernel_size=(3, 3), stride=(1, 1), padding=(1, 1))
        )
        (3): ConvModule(
          (conv): Conv2d(256, 256, kernel_size=(3, 3), stride=(1, 1), padding=(1, 1))
        )
      )
    )
    (rpn_head): RPNHead(
      (loss_cls): CrossEntropyLoss()
      (loss_bbox): SmoothL1Loss()
      (rpn_conv): Conv2d(256, 256, kernel_size=(3, 3), stride=(1, 1), padding=(1, 1))
      (rpn_cls): Conv2d(256, 12, kernel_size=(1, 1), stride=(1, 1))
      (rpn_reg): Conv2d(256, 48, kernel_size=(1, 1), stride=(1, 1))
    )
    (bbox_roi_extractor): SingleRoIExtractor(
      (roi_layers): ModuleList(
        (0): RoIAlign(out_size=(7, 7), spatial_scale=0.25, sample_num=2, use_torchvision=False)
        (1): RoIAlign(out_size=(7, 7), spatial_scale=0.125, sample_num=2, use_torchvision=False)
        (2): RoIAlign(out_size=(7, 7), spatial_scale=0.0625, sample_num=2, use_torchvision=False)
        (3): RoIAlign(out_size=(7, 7), spatial_scale=0.03125, sample_num=2, use_torchvision=False)
      )
    )
    (bbox_head): SharedFCBBoxHead(
      (loss_cls): CrossEntropyLoss()
      (loss_bbox): SmoothL1Loss()
      (fc_cls): Linear(in_features=1024, out_features=2, bias=True)
      (fc_reg): Linear(in_features=1024, out_features=8, bias=True)
      (shared_convs): ModuleList()
      (shared_fcs): ModuleList(
        (0): Linear(in_features=12544, out_features=1024, bias=True)
        (1): Linear(in_features=1024, out_features=1024, bias=True)
      )
      (cls_convs): ModuleList()
      (cls_fcs): ModuleList()
      (reg_convs): ModuleList()
      (reg_fcs): ModuleList()
      (relu): ReLU(inplace=True)
    )
}>
```